# Modification of coupled integral equations method for calculation the accelerating structure characteristics

M.I.Ayzatsky*

National Science Center

Kharkov Institute of Physics and Technology (NSC KIPT),

610108, Kharkov, Ukraine

**Abstract**

Modification of coupled integral equations method (CIEM) for calculating the characteristics of the accelerating structures is presented. In earlier developed CIEM schemes the coupled integral equations are derived for the unknown electrical fields at interfaces that divide the adjacent volumes. In addition to the standard division of the structured waveguide by interfaces between the adjacent cells, we propose to introduce new interfaces in places where electric field has the simplest transverse structure. Moreover, the system of coupled integral equations is formulated for longitudinal electrical fields in contrast to the standard approach where the transverse electrical fields are unknowns. The final matrix equations contain expansion coefficients of the longitudinal electric field at these additional interfaces. This modification makes it possible to deal with a physical quantity that plays an important role in the acceleration of particles (a longitudinal electric field), and to obtain approximate equations for the case of a slow change in the waveguide parameters

## 1 Introduction

The main characteristic of the slow-wave accelerating structures is the distribution of the electric field in both steady state and transient modes. This imposes certain restrictions on the methods of calculating their characteristics, manufacturing and tuning. The slow-wave accelerating structures mainly belong to the class of structured waveguides - waveguides that consist of similar, but not always identical, cells ( disk-loaded waveguides, chains of coupled resonators, etc.). One of the effective approaches for calculating the characteristics of structured waveguides is the coupled integral equations method (CIEM) [1, 2, 3, 4]. There are only a few works in which this approach was used to study accelerating structures [5, 6]. There are other methods that can be used for calculation of characteristics of structured waveguides (see, for example, [7, 8, 9, 10]). Some studies of characteristics of accelerator structures are based on these methods [11, 12] The advantage of CIEM is the formulation of the second order difference equation, on the basis of which a numerical algorithm is developed. The presence of an equation makes it possible not only to perform calculations, but also to analyse and develop simplified models.

---

*aizatsky@kipt.kharkov.ua

Based on a system of coupled integral equations, an approximate method is constructed for calculating the characteristics of structured waveguides with slowly varying dimensions [5] . It is the analog of classical Eikonal and WKB methods with taking into account not only propagating waves, but also evanescent ones. The advantage of this approach is the simple physical (but not simple mathematical) interpretation of obtained equations and their solutions. This approximate method was used to study the characteristics of the simplest case of structured waveguide — a Disk-Loaded Waveguide (DLW) with very thin diaphragms [6].

Analysis of the standard method of coupled integral equations for studying the characteristics of DLS with real geometry showed that some modifications of the standard approach can be useful. In this paper we present such modification of coupled integral equations method for calculating the characteristics of the accelerating structures. In earlier developed CIEM schemes the coupled integral equations are derived for the unknown electrical fields at interfaces that divide the adjacent volumes. Usually these interfaces include geometrical singularities, such as sharp edges. In this case it is needed to use special basis functions. In addition to the standard division of the structured waveguide by interfaces between the adjacent cells, we propose to introduce new interfaces in places where electric field has the simplest transverse structure. Moreover, the system of coupled integral equations is formulated for longitudinal electrical fields in contrast to the standard approach where the transverse electrical fields are unknowns. The final matrix equations contain expansion coefficients of the longitudinal electric field at these additional interfaces. This modification makes it possible to deal with a physical quantity that plays an important role in the acceleration of particles (a longitudinal electric field).

In the new formulation, the method of coupled integral equations becomes not only a numerical tool, but also the important part of the mathematical procedure for obtaining approximate equations describing structured waveguides with a slow change in its parameters. Moreover, media with local periodicity is a very attractive area of modern wave physics (see, for example, [13]) and proposed approach can be useful for new applications.

## 2  Accelerating structure model. Basic equations

Consider a segment of DLW (circular corrugated waveguide), the geometry of which is shown in the Figure 1.

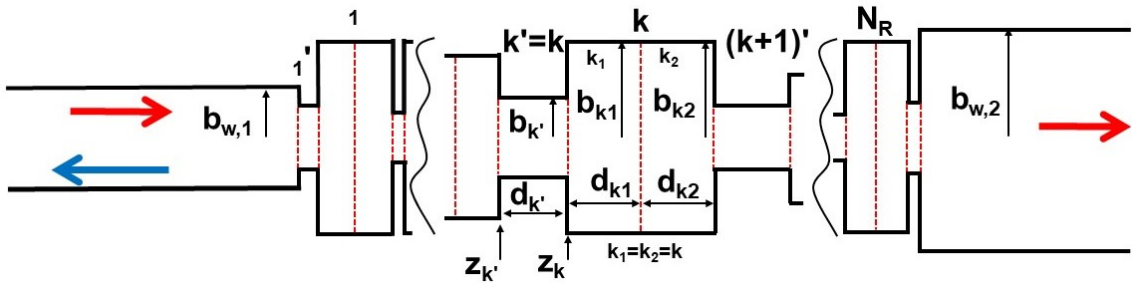

Figure 1: Chain of pieces of cylindrical waveguides that is connected with semi-infinite cylindrical waveguides

All segment volumes are filled with dielectric ($\varepsilon = \varepsilon' + i\varepsilon''$, $\varepsilon'' \geq 0$). We divide the DLW into subdomains, each of which is a circular waveguide. Unlike earlier works [14, 1, 2, 3, 4, 11, 12, 5, 6], we divide each volume with large cross-section into two equal subvolumes (in general, they can be different). Volumes with large cross section will be numbered by the index $k (1 \leq k \leq N_{REZ})$, subvolumes by $k_1$ and $k_2$ ($k_1 = k_2 = k0$. A small cross-sectional volume, placed to the left of a large cross-sectional volume with an index $k$, will be numbered by the index $k'$ ($1 \leq k' \leq (N_{REZ} + 1)'$).

We will consider only axially symmetric TM fields with $E_z, E_r, H_\varphi$ components. Time dependence is $\exp(-i\omega t)$. Since we are interested in considering accelerating structures, we must remember that it will be necessary to take into account the beam loading. Therefore, we will use initial expansions that are slightly different from the standard CIEM approach and give the possibility to include current into consideration. In each cylindrical volume (with index $q$) we expand the electromagnetic field electromagnetic field in terms of the complete orthogonal set of transverse functions

$$E_z^{(q)}(r, z_q + \tilde{z}) = \sum_m E_{z,m}^{(q)}(\tilde{z}) J_0\left(\frac{\lambda_m}{b_q} r\right),$$

$$E_r^{(q)}(r, z_q + \tilde{z}) = \sum_m E_{r,m}^{(q)}(\tilde{z}) J_1\left(\frac{\lambda_m}{b_q} r\right), \quad (1)$$

$$H_\varphi^{(q)}(r, z_q + \tilde{z}) = \sum_m H_{\varphi,m}^{(q)}(\tilde{z}) J_1\left(\frac{\lambda_m}{b_q} r\right),$$

where $0 \leq \tilde{z} \leq d_q, \gamma_m^{(q)2} = \left(\frac{\lambda_m}{b_q}\right)^2 - \frac{\omega^2}{c^2}\varepsilon$, $\operatorname{Im}\gamma_m^{(q)} > 0$, $\operatorname{Re}\gamma_m^{(q)} < 0$, $\gamma_{-m}^{(q)} = -\gamma_m^{(q)}$, $J_0(\lambda_m) = 0$ From Maxwell equations we obtain

$$\frac{d^2 E_{r,m}^{(q)}}{d\tilde{z}^2} - \gamma_m^{(q)2} E_{r,m}^{(q)} = -\gamma_m^{(q)2}\frac{1}{i\omega\varepsilon_0\varepsilon} I_{r,m}^{(q)} - \frac{1}{i\omega\varepsilon_0\varepsilon}\frac{\lambda_m}{b_q}\frac{dI_{z,m}^{(q)}}{dz} = F_m^{(q)}(\tilde{z}) \quad (2)$$

$$H_{\varphi,m}^{(q)} = \frac{1}{\gamma_m^{(q)2}}\left(\frac{\lambda_m}{b_k} I_{z,m}^{(q)} + i\omega\varepsilon_0\varepsilon\frac{dE_{r,m}^{(q)}}{d\tilde{z}}\right),$$

$$E_{z,m}^{(q)} = -\frac{\lambda_m}{b_k}\frac{1}{\gamma_m^{(q)2}}\frac{dE_{r,m}^{(q)}}{d\tilde{z}} + \frac{i\omega}{\varepsilon_0 c^2 \gamma_m^{(q)2}} I_{z,m}^{(q)}, \quad (3)$$

$$I_{r,m}^{(k)}(\tilde{z}) = \frac{1}{W_m^{(k)}} \int_0^{2\pi}\int_0^{b_k} j_r(r, \varphi, z_k + \tilde{z}) J_1\left(\frac{\lambda_m}{b_k} r\right) r\, dr\, d\varphi,$$

$$I_{z,m}^{(k)}(\tilde{z}) = \frac{1}{W_m^{(k)}} \int_0^{2\pi}\int_0^{b_k} j_z(r, \varphi, z_k + \tilde{z}) J_0\left(\frac{\lambda_m}{b_k} r\right) r\, dr\, d\varphi,$$

$$W_m^{(k)} = \pi b_k^2 J_1^2(\lambda_m).$$

The system of equations (2)-(3) is the base for the study of electromagnetic fields in accelerating sections.

In the semi-infinite waveguides the electromagnetic field can be expanded in terms of the TM eigenmodes $\vec{\mathcal{E}}_s^{(w,p)}, \vec{\mathcal{H}}_s^{(w,p)}$ of a circular waveguide ($p = 1, 2$)

$$\vec{H}^{(w,p)} = \sum_s \left(G_s^{(p)} \vec{\mathcal{H}}_s^{(w,p)} + G_{-s}^{(k)} \vec{\mathcal{H}}_{-s}^{(w,p)}\right)$$

$$\vec{E}^{(w,p)} = \sum_s \left(G_s^{(p)} \vec{\mathcal{E}}_s^{(w,p)} + G_s^{(-p)} \vec{\mathcal{E}}_{-s}^{(w,p)}\right) \quad (4)$$

On the introduced interfaces we represent the electric fields as series of basis functions

$$E_r^{(k')}(r, d_{k'}) = \sum_s C_s^{(k_1)} \varphi_s^{(r)}(r/b_{k'})$$
$$E_r^{((k+1)')}(r, 0) = \sum_s C_s^{(k_2)} \varphi_s^{(r)}(r/b_{(k+1)'})$$
(5)

$$E_z^{(k_1)}(r, d_k/2) = E_z^{(k_2)}(r, 0) = \sum_s Q_s^{(k)} \varphi_s^{(z)}(r/b_k) \qquad (6)$$

The boundary conditions for electric fields at the junctions are written as

$$\sum_m E_{r,m}^{(k')}(d_{k'}) J_1\left(\frac{\lambda_m}{b_{k'}}r\right) = \sum_s C_s^{(k_1)} \varphi_s^{(r)}(r/b_{k'}), 0 \leqslant r < b_{k'},$$

$$\sum_m E_{r,m}^{(k_1)}(0) J_1\left(\frac{\lambda_m}{b_k}r\right) = \begin{cases} \sum_s C_s^{(k_1)} \varphi_s^{(r)}(r/b_{k'}), 0 \leqslant r < b_{k'}, \\ 0, \qquad b_{k'} \leqslant r < b_k, \end{cases}$$

$$\sum_m E_{z,m}^{(k_1)}(d_k/2) J_0\left(\frac{\lambda_m}{b_k}r\right) = \sum_m E_{z,m}^{(k_2)}(0) J_0\left(\frac{\lambda_m}{b_k}r\right) =$$
$$\sum_s Q_s^{(k)} \varphi_s^{(z)}(r/b_k), 0 \leqslant r < b_k,$$
(7)

$$\sum_m E_{r,m}^{(k_2)}(d_k/2) J_1\left(\frac{\lambda_m}{b_k}r\right) = \begin{cases} \sum_s C_s^{(k_2)} \varphi_s^{(r)}(r/b_{(k+1)'}), 0 \leqslant r < b_{(k+1)'}, \\ 0, \qquad b_{(k+1)'} \leqslant r < b_k, \end{cases}$$

$$\sum_m E_{r,m}^{(k+1)'}(0) J_1\left(\frac{\lambda_m}{b_{(k+1)'}}r\right) = \sum_s C_s^{(k_2)} \varphi_s^{(r)}(r/b_{(k+1)'}), 0 \leqslant r < b_{(k+1)'}.$$

Using the completeness and orthogonality of Bessel functions $J_0(r\lambda_m/b)$ and $J_1(r\lambda_m/b)$, it is easy to find from (7) coefficients of the left series. It should be noted that that the boundary conditions (7) contain also the longitudinal electric fields.

In the standard CIEM approach, the second group of boundary conditions contains, as a rule, the continuity of the tangential components of the magnetic field

$$\sum_m H_{\varphi,m}^{(k')}(d_{k'}) J_1\left(\frac{\lambda_m}{b_{k'}}r\right) = \sum_m H_{\varphi,m}^{(k_1)}(0) J_1\left(\frac{\lambda_m}{b_k}r\right), \qquad 0 < r < b_{k'},$$
$$\sum_m H_{\varphi,m}^{(k_2)}(d_k/2) J_1\left(\frac{\lambda_m}{b_k}r\right) = \sum_m H_{\varphi,m}^{(k'+1)}(0) J_1\left(\frac{\lambda_m}{b_{(k+1)'}}r\right), \quad 0 < r < b_{(k+1)'},$$
(8)

Multiplying the right and left sides of this relations by a testing functions $\psi_{s'}(r/b_{k'})$ and integrating with respect to $r$ from 0 to $b_{k'}$, we get such equations

$$\sum_m H_{\varphi,m}^{(k')}(d_{k'}) R_{s',m}^{\psi(k',k')} = \sum_m H_{\varphi,m}^{(k_1)}(0) R_{s',m}^{\psi(k',k)},$$
$$\sum_m H_{\varphi,m}^{(k_2)}(d_k/2) R_{s',m}^{\psi((k+1)',k)} = \sum_m H_{\varphi,m}^{(k+1)'}(0) R_{s',m}^{\psi((k+1)',(k+1)')}.$$
(9)

In our case it is needed to add additional conditions the continuity of the electric field tangential components at the interfaces in the middle of large cross-sectional volumes

$$\sum_m E_{r,m}^{(k_1)}(d_k/2) J_1\left(\frac{\lambda_m}{b_k}r\right) = \sum_m E_{r,m}^{(k_2)}(0) J_1\left(\frac{\lambda_m}{b_k}r\right) \Rightarrow E_{r,m}^{(k_1)}(d_k/2) = E_{r,m}^{(k_2)}(0) \quad (10)$$

We will consider the case when the dimensions of two semi-infinite waveguides are chosen such that only the dominant mode $TM_{01}$ propagates, and the higher-order modes are all evanescent We will suppose that there is an incident wave that travels from $z = -\infty$ with amplitude $G_1^{(1)} = 1$ ($G_n^{(1)} = 0, n \geqslant 2$). Using the standard CIEM technique, we obtain such systems of vector equations $k = 1, \ldots, N_{REZ}$ (see Appendix 1)

$$\begin{aligned}
-T'^{(1,k')}C^{(k_2-1)} + \left(T'^{(2,k')} + T^{(2,k',k)}\right)C^{(k_1)} + T^{(1,k',k)}Q^{(k)} &= Z^{-(k)}, \\
T^{(1,k',k)}Q^{(k)} - \left(T'^{(2,(k+1)')} + T^{(2,(k+1)',k)}\right)C^{(k_2)} + T'^{(1,(k+1)')}C^{((k+1)_1)} &= Z^{+(k)}, \quad (11)\\
T^{r(k'+1,k)}C^{(k_2)} - T^{r(k',k)}C^{(k_1)} + T^{z(k)}Q^{(k)} &= Z^{(k)},
\end{aligned}$$

and two additional equations

$$\begin{aligned}
-\varepsilon T'^{(1,1')}C^{(1_1)} + \varepsilon T'^{(2,1')}C^{(L)} - T^{(L)}C^{(L)} &= R^{(L)} + Z^{(L)}, \\
T^{(R)}C^{(R)} - \varepsilon T'^{(2,(N_R+1)')}C^{(R)} + \varepsilon T'^{(1,(N_R+1)')}C^{((N_{REZ})_2)} &= Z^{(R)},
\end{aligned} \quad (12)$$

where $C^{(L)} = C^{(0_2)}$ and $C^{(R)} = C^{((N_{REZ}+1)_1)}$ are the vectors of expansion coefficients of the tangential components of the electric field on the left and right interfaces of the DLW and the semi-infinite waveguides, $Z$ (with different superscripts) are "current" vectors (see Appendix 1) that equal zero in the absence of current, $T$ (with different superscripts) are the matrices defined in Appendix 1.

For the numerical solution of system (11), it is necessary to limit the number of basis and testing functions $\varphi_s(r), \varphi_s(z), \psi_s$. We will suppose that $\varphi_s(r) = \psi_s(r) \equiv 0, s > N_R, \varphi_s(z) \equiv 0, s > N_Z$, therefore $C^{(k)} \in \mathbb{C}^{N_R}, Q^{(k)} \in \mathbb{C}^{N_Z}$.

Then we will have such sizes of defined matrices: $T'^{(1,k')}, T'^{(2,k')}, T^{(2,k',k)} \in \mathbb{C}^{N_\mathbb{R} \times N_\mathbb{R}}, T^{(1,k',k)} \in \mathbb{C}^{N_\mathbb{R} \times N_\mathbb{Z}}, T^{r(k',k)} \in \mathbb{C}^{N_\mathbb{Z} \times N_\mathbb{R}}$ and $T^{z(kk)} \in \mathbb{C}^{N_\mathbb{Z} \times N_\mathbb{Z}}$.

Amplitudes of the eigen waves in the semi-infinite waveguides are determined by the expansion coefficients $C^{(L)}$ and $C^{(R)}$

$$\begin{aligned}
G_{-1}^{(1)} &= 1 + 2\frac{b_{1'}^2 \lambda_1}{J_1^2(\lambda_1) b_{w_1}^2 \gamma_1^{(w_1)} b_{w_1}} \sum_{s'} R_{1,s'}^{w,L} C_{s'}^{(L)} \\
G_{-s}^{(1)} &= -2\frac{b_{1'}^2 \lambda_s}{J_1^2(\lambda_s) b_{w_1}^2 \gamma_{-s}^{(w_1)} b_{w_1}} \sum_{s'} R_{s,s'}^{w,L} C_{s'}^{(L)}, \quad s = 2,3,... \quad (13)\\
G_s^{(2)} &= -2\frac{b_{(N_R+1)'}^2 \lambda_s}{J_1^2(\lambda_s) b_{w_2}^2 \gamma_s^{(w_2)} b_{w_2}} \sum_{s'} R_{s,s'}^{w,R} C_{s'}^{(R)}, \quad s = 1,2,...
\end{aligned}$$

where $\gamma_s^{(w,p)2} = \lambda_s^2/b_{w,p}^2 - \omega^2/c^2$, $R_{m,s}^{w,L} = \int_0^1 \varphi_s^{(r)}(x) J_1(b_{1'}\lambda_m x/b_{w_1}) x dx$, and

$R_{m,s}^{w,R} = \int_0^1 \varphi_s^{(r)}(x) J_1(b_{(N_R+1)'}\lambda_m x/b_{w_2}) x dx$.

# 3 Infinitive uniform disk loaded waveguide

To demonstrate the difference between the standard and the proposed approaches, consider an infinite homogeneous disk-loaded waveguide without current $b_{k'} = a, \ d_{k'} = t, \ b_k = b, \ d_k = d$.

If we omit the presence of boundaries for the uniform segment, we obtain from (11) the equations that describe such waveguide. These difference equations in the matrix form are written as

$$\begin{cases} \left(T'^{(2)} + T^{(2)}\right) C^{(k_1)} = T'^{(1)} C^{(k_2-1)} - T^{(1)} Q^{(k)}, \\ \left(T'^{(2)} + T^{(2)}\right) C^{(k_2)} = T'^{(1)} C^{(k_1+1)} + T^{(1)} Q^{(k)}, \\ T^r C^{(k_2)} - T^r C^{(k_1)} + T^z Q^{(k)} = 0, \end{cases} \quad (14)$$

We supposed that matrices $T'^{(1)}, T'^{(2)}, T^{(2)}, T^z$ are invertible.

Excluding $C^{(k_2)}$ and $Q^{(k)}$ from (14), we get the standard matrix difference equation

$$\bar{T} C^{(k_1)} = \bar{T}^{(+)} C^{(k_1+1)} + \bar{T}^{(-)} C^{(k_1-1)} \quad (15)$$

where

$$\begin{aligned} \bar{T} &= \bar{\bar{T}} - T'^{(1)} \bar{\bar{T}}^{-1} T'^{(1)} - T^{(1)} T^{z-1} T^r \bar{\bar{T}}^{-1} T^{(1)} T^{z-1} T^r, \\ \bar{T}^{(-)} &= \left(T'^{(1)} \bar{\bar{T}}^{-1}\right) T^{(1)} T^{z-1} T^r, \\ \bar{T}^{(+)} &= T^{(1)} T^{z-1} T^r \left(\bar{\bar{T}}^{-1} T'^{(1)}\right), \\ \bar{\bar{T}} &= T'^{(2)} + T^{(2)} + T^{(1)} T^{z-1} T^r. \end{aligned} \quad (16)$$

The size of matrices $\bar{T}, \bar{T}^{(+)}, \bar{T}^{(-)} \in \mathbb{C}^{N_R \times N_R}$ is defined by the number of basis functions $\varphi_s^{(r)}$ in the $E_r$ expansion. The $\bar{T}^{(\pm)}$ matrices are the product of two square matrices, one of which is the product of the rectangular matrices $\left(T^{(1)} T^{z-1} T^r\right)$. As $T^{z-1} T^r \in \mathbb{C}^{N_Z \times N_R}$ and $\left(T^{(1)}\right) \in \mathbb{C}^{N_R \times N_Z}$, then for $N_R \leqslant N_Z$ it follows from Binet-Cauchy formula that $\det \left(T^{(1)} T^{z-1} T^r\right) \neq 0$ and we can inverse the matrix $\bar{T}^{(+)}$ and get an equation of the form [2]

$$Y^{(k+1)} = H Y^{(k)}, \quad (17)$$

where $H$ is a block companion matrix

$$H = \begin{pmatrix} \bar{T}^{(+)-1} \bar{T} & -\bar{T}^{(+)-1} \bar{T}^{(-)} \\ I & 0 \end{pmatrix} \quad (18)$$

and

$$Y^{(k)} = \begin{pmatrix} C^{(k+1)} \\ C^{(k)} \end{pmatrix}$$

It can be shown that for basis $\varphi_s^{(z)} = J_0(\lambda_s x), \ x \in [0, 1]$

$$\begin{aligned} \left(T^{(1)} T^{z-1} T^r\right)_{s',s} &= \sum_{m=1}^{N_Z} T_{s',m}^{(1)} \left(T^{z-1} T^r\right)_{m,s} = \\ &= \frac{b_k}{b_*} \frac{b_{k'}^2}{b_k^2} \sum_{m=1}^{N_Z} \frac{2}{J_1^2(\lambda_m)} \frac{1}{b_k \gamma_m^{(k)} sh\left(\gamma_m^{(k)} d_k\right)} R_{s',m}^{\psi(k',k)} R_{m,s}^{\varphi,r(k',k)}, \ s',s = [\![1, N_R]\!], \end{aligned} \quad (19)$$

Since $N_z$ appears only as an upper summation limit in (19) and is not included in the final equation (15), it can be chosen large enough without limiting the size of the vectors $C^{(k_1)}$.

Difference equation (15) is not symmetric ( $\bar{T}^{(+)} \neq \bar{T}^{(-)}$ ) since it includes only vectors describing the fields on the left side of volumes with a large cross section. These fields have a different interaction with right and left neighbors. The absence of symmetry makes it more difficult [1] to apply a transformation [15, 16], which gives simple method of finding Floquet coefficients and possibility to use the WKB approach.

Eliminating $C^{(k_1)}$ and $C^{(k_2)}$, we can transform (14) into a symmetric difference equation

$$\tilde{T} Q^{(k)} = \tilde{T}^{(+)} Q^{(k+1)} + \tilde{T}^{(-)} Q^{(k-1)}, \tag{20}$$

where

$$\tilde{T} = T^z + 2T^r \left\{ \left( T'^{(2)} + T^{(2)} \right) - T'^{(1)} \left( T'^{(2)} + T^{(2)} \right)^{-1} T'^{(1)} \right\}^{-1}, \tag{21}$$

$$\tilde{T}^{(+)} = \tilde{T}^{(-)} = T^r \tilde{\tilde{T}} T^{(1)}, \tag{22}$$

$$\tilde{\tilde{T}} = \left\{ \left( T'^{(2)} + T^{(2)} \right) - T'^{(1)} \left( T'^{(2)} + T^{(2)} \right)^{-1} T'^{(1)} \right\}^{-1} T'^{(1)} \left( T'^{(2)} + T^{(2)} \right)^{-1}. \tag{23}$$

Matrices $T^r \in \mathbb{C}^{N_Z \times N_R}$ and $\left( \tilde{\tilde{T}} T^{(1)} \right) \in \mathbb{C}^{N_R \times N_Z}$, then for $N_Z \leq N_R$ it follows from Binet-Cauchy formula that $\det \left( T^r \tilde{\tilde{T}} T^{(1)} \right) \neq 0$ and we can inverse the matrix $\tilde{T}^{(+)}$ and get an equation of the form $\left( T = \tilde{T}^{(+)-1} \tilde{T} \right)$

$$T Q^{(k)} = Q^{(k+1)} + Q^{(k-1)}, \tag{24}$$

which is the basis for using the solution decomposition method [15]. Condition $N_Z \leq N_R$ is the exact opposite of the condition that is required to find solutions to equation (15).

The size of matrix $\tilde{T} \in \mathbb{C}^{N_Z \times N_Z}$ is defined by the number of basis functions $\varphi_s^{(z)}(r/b_k)$ in the $E_z$ expansion. The $E_r$ expansion contains $N_R$ basis functions $\varphi_s^{(r)}$. As $N_R \geqslant N_Z$, we can improve the accuracy of $E_r$ representation (to increase $N_R$) without increasing the size of matrix $T$ ($N_Z \times N_Z$). It should also be noted that matrix $T$ is not Hermitian.

Using the transformation [15]

$$\begin{aligned} Q^{(k)} &= Q^{(k,1)} + Q^{(k,2)} \\ Q^{(k+1)} &= M^{(1)} Q^{(k,1)} + M^{(2)} Q^{(k,2)} \end{aligned} \tag{25}$$

where

$$\left( T M^{(i)} - M^{(i)2} - I \right) = 0, \tag{26}$$

we get ($i = 1, 2$)

$$Q^{(k+1,i)} = M^{(i)} Q^{(k,i)}. \tag{27}$$

It can be shown that for $N_Z \leqslant N_R$ the matrix $T$ is non-defective [2], and can be decomposed as

$$T = U \Theta U^{-1}, \tag{28}$$

where $U$ is the matrix of eigen vectors $U_s$ and $\Theta = diag(\theta_1, \theta_2, ...)$, $\theta_s$ - eigen values.

---

[1] Matrix equations, whose solutions are necessary to construct the WKB equations, become more complicated

[2] The infinitive uniform disk-loaded waveguide has $2 \times N_Z$ different independent solutions (waves)

Then the solutions of quadratic matrix equations (26) are $i = 1, 2$

$$M^{(i)} = U\Lambda^{(i)}U^{-1}, \qquad (29)$$

where $\Lambda^{(i)} = diag(\lambda_1^{(i)}, \lambda_2^{(i)}, ...)$ and $\lambda_s^{(i)}$ are the solutions of the characteristic equations

$$\begin{aligned}\lambda_s^{(i)2} - \theta_s \lambda_s^{(i)} + 1 &= 0, \\ \lambda_s^{(1)} &= \theta_s/2 + \sqrt{(\theta_s/2)^2 - 1}, \\ \lambda_s^{(2)} &= \theta_s/2 - \sqrt{(\theta_s/2)^2 - 1}.\end{aligned} \qquad (30)$$

The matrices $M^{(i)}$ have the same eigen vectors, therefore they are commutative. As $\lambda_s^{(1)}\lambda_s^{(2)} = 1$, the matrices $M^{(i)}$ satisfy the condition $M^{(1)}M^{(2)} = I$. We will suppose that $\left|\text{Re}\left(\lambda_s^{(1)}\right)\right| < 1$ and $\left|\text{Re}\left(\lambda_s^{(2)}\right)\right| > 1$.

Representing the vector $Q^{(k)}$ as the sum of two new vectors $Q^{(k,1)}$ and $Q^{(k,2)}$ we did not assume that they are individually solutions to the difference equation (24). Let us show that when $M^{(i)}$ are chosen as solutions to (26), the vectors $Q^{(k,1)}$ and $Q^{(k,2)}$ are independent solutions to the equation (24).

If we know the radial distribution of longitudinal components of electric fields in two consecutive sections of the waveguide $Q^{(0)}$, $Q^{(1)}$, then we can find vectors $Q^{(0,1)}$, $Q^{(0,2)}$

$$\begin{aligned}Q^{(0,1)} &= \left(M^{(2)} - M^{(1)}\right)^{-1} \left(M^{(2)}Q^{(0)} - Q^{(1)}\right) \\ Q^{(0,2)} &= -\left(M^{(2)} - M^{(1)}\right)^{-1} \left(M^{(1)}Q^{(0)} - Q^{(1)}\right)\end{aligned} \qquad (31)$$

To find the solutions of equations (27) with conditions (31) and the conditions at the infinity for all values of $k$ we have to consider the equations (27) for $k > 0$ and $k < 0$ separately. Then the solutions of the difference matrix equations (27) with taking into account the conditions at the infinity are

$$\begin{aligned}Q^{(k,1)} &= M^{(1)k}Q^{(0,1)}, \quad k \geqslant 0, \\ Q^{(k,2)} &= M^{(2)k}Q^{(0,2)}, \quad k \leqslant 1.\end{aligned} \qquad (32)$$

Vectors $Q^{(0)}$ and $Q^{(1)}$ we can represent as a sum of eigen vectors ($i = 0, 1$)

$$Q^{(i)} = \sum_s A_s^{(i)} U_s. \qquad (33)$$

The matrix $T$ is not Hermitian and the vectors $U_s$ are not orthogonal. In this case

$$A_s^{(i)} = \sum_{s'} \left(U^{-1}\right)_{s,s'} Q_{s'}^{(i)}. \qquad (34)$$

Substitution (34) into (31) gives

$$\begin{aligned}Q^{(0,1)} &= \left(M^{(2)} - M^{(1)}\right)^{-1} \sum_s \left(\lambda_s^{(2)} A_s^{(0)} - A_s^{(1)}\right) U_s = \sum_s \frac{\left(\lambda_s^{(2)} A_s^{(0)} - A_s^{(1)}\right)}{\lambda_s^{(2)} - \lambda_s^{(1)}} U_s, \\ Q^{(0,2)} &= -\left(M^{(2)} - M^{(1)}\right)^{-1} \sum_s \left(\lambda_s^{(1)} A_s^{(0)} - A_s^{(1)}\right) U_s = -\sum_s \frac{\left(\lambda_s^{(1)} A_s^{(0)} - A_s^{(1)}\right)}{\lambda_s^{(2)} - \lambda_s^{(1)}} U_s.\end{aligned} \qquad (35)$$

Then the solution of the equation (25) takes the form

$$Q^{(k)} = \begin{cases} -\sum_s \dfrac{\lambda_s^{(2)k}\left(\lambda_s^{(1)}A_s^{(0)} - A_s^{(1)}\right)}{\lambda_s^{(2)} - \lambda_s^{(1)}} U_s, & k < -1. \\[2mm] \sum_s \left\{ \dfrac{\lambda_s^{(1)k}\left(\lambda_s^{(2)}A_s^{(0)} - A_s^{(1)}\right)}{\lambda_s^{(2)} - \lambda_s^{(1)}} - \dfrac{\lambda_s^{(2)k}\left(\lambda_s^{(1)}A_s^{(0)} - A_s^{(1)}\right)}{\lambda_s^{(2)} - \lambda_s^{(1)}} \right\} U_s, & k = 0, 1, \\[2mm] \sum_s \dfrac{\lambda_s^{(1)k}\left(\lambda_s^{(2)}A_s^{(0)} - A_s^{(1)}\right)}{\lambda_s^{(2)} - \lambda_s^{(1)}} U_s, & k > 1. \end{cases} \quad (36)$$

For the case when $Q^{(0)} = U_m$ and $Q^{(1)} = \lambda_m^{(1)}U_m$ we have $A_s^{(0)} = \delta_{s,m}$, $A_s^{(1)} = \lambda_m^{(1)}\delta_{s,m}$ and

$$Q^{fw(k)} = \begin{cases} 0, & k < 0 \\ \lambda_m^{(1)k}U_m, & k \geqslant 0 \end{cases} \quad (37)$$

For the case $Q^{(1)} = \lambda_m^{(2)}U_m$

$$Q^{bw(k)} = \begin{cases} \lambda_m^{(2)k}U_m, & k \leqslant 1, \\ 0, & k > 1. \end{cases} \quad (38)$$

Therefore, the vector sequences $\lambda_s^{(i)k}U_s$ can be considered as forward ($i=1$) or backward ($i=2$) eigen solutions of the equation (24).

It was shown [5] that the vector equation (27) can be transformed into a difference equation for any component of the vector $Q^{(k,i)}$. For a homogeneous waveguide these equations have the same form. Therefore, if we choose basis function that fulfill a condition $\varphi_s^{(z)}(0) = 0$ (for example, $J_0(\lambda_s r/b)$), we can write a difference equation that connects the values of electric field $E_z^{(k)} = \sum_s \left(Q_s^{(k,1)} + Q_s^{(k,2)}\right)$ at different points on the axis $r=0$, $z_k = k(d+t) + d/2$

$$\widehat{det} \begin{pmatrix} \widehat{L}_1 & -T_{1,2} & \ldots & -T_{1,N_z} \\ -T_{2,1} & \widehat{L}_2 & \ldots & \ldots \\ \ldots & \ldots & \ldots & \ldots \\ -T_{N_z,1} & -T_{N_z,2} & \ldots & \widehat{L}_{N_z} \end{pmatrix} E_z^{(k)} = 0, \quad (39)$$

where the operator $\widehat{det}$ is defined on the base of rules of common determinants with special order of multiplication

$$\widehat{det}\begin{pmatrix} \widehat{L}_1 & -T_{1,2} \\ -T_{2,1} & \widehat{L}_2 \end{pmatrix} = \widehat{L}_1\widehat{L}_2 - T_{1,2}T_{2,1}, \quad (40)$$

$\widehat{L}_i = \widehat{\sigma}^+ + \widehat{\sigma}^- - T_{i,i}$, $\widehat{\sigma}^+ \left(\widehat{\sigma}^+ b^{(k)} = b^{(k+1)}\right)$ and $\widehat{\sigma}^- \left(\widehat{\sigma}^- b^{(k)} = b^{(k-1)}\right)$ are shift operators. It was shown that equation (39) does not have spurious solutions as it was for the equation obtained on the basis of a coupled cavities model [17].

# 4 Modified Vector Equations

The system of vector equations (11) can be transformed to a system with only unknowns $Q^{(k)}$ (see Appendix 2)

$$T^{(Q_1)}Q^{(1)} + T^{(Q_2)}Q^{(2)} = Z^{Q(1)},$$
$$T^{(k)}Q^{(k)} = T^{+(k)}Q^{(k+1)} + T^{-(k)}Q^{(k-1)} + Z^{Q(k)}, \quad k = 2, ..., N_{REZ} - 1, \quad (41)$$
$$T^{(Q_{N_{REZ}}-1)}Q^{(N_{REZ}-1)} + T^{(Q_{N_{REZ}})}Q^{(N_{REZ})} = Z^{Q(N_{REZ})},$$

where the sizes of all $T$ matrices are $N_Z \times N_Z$. There are additional equations relating $Q^{(k)}$, $C^{(L)}$, $C^{(R)}$, from which we can calculate the reflection and transmission coefficients (13) (see Appendix 2).

System (41) is similar to that analyzed in [5] and, therefore, can be the basis for obtaining the WKB equations.

A mathematical model of the accelerating structure must satisfy the following criteria:

- a model must provide the necessary accuracy for calculating both electromagnetic fields and fulfillment of the law of conservation of energy;

- for a homogeneous case it must give correct dispersion characteristics (phase shift per cell, phase and group velocity).

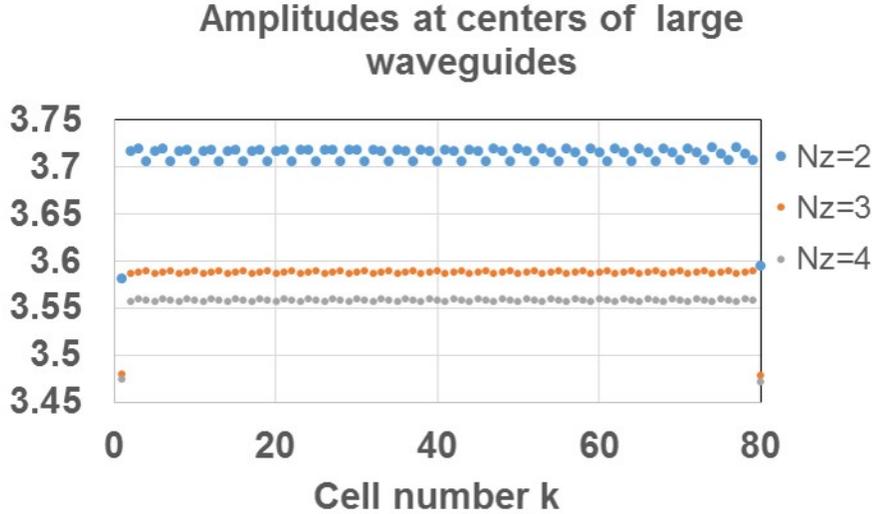

Figure 2: Amplitudes of logitudinal electric fields $E_z$ at centers of large waveguides ($r = 0$, $\tilde{z} = d_{k_1}$) for the case of homogeneous structure

We have developed a code on the base of FORTRAN that solves the system (41). Results of calculations for the homogeneous structured waveguide ($N_{REZ} = 80$, $\varepsilon = 1$) with such geometry: $b_{k'} = a = 1.3$ cm, $d_{k'} = t = 0.4$ cm, $b_k = b = 4.1409$ cm, $d_k = d = (3.4989\text{-}0.4)/2$cm, $b_{1'} = b_{(N_{REZ}+1)'} = 1.8632$cm, $b_1 = b_{N_{REZ}} = 4.2025$cm are presented in Figure 2, Figure 3 and Table 1. At these parameters the reflection and transition coefficient are equal $R = 6.3\text{E-}04$ and $T = 0.9999$. Power flows through all interfaces are constants and equal to the input power flow to the 4-th decimal place ( $P = 2.5292$).Thus, the law of conservation of energy is satisfied with sufficient accuracy.

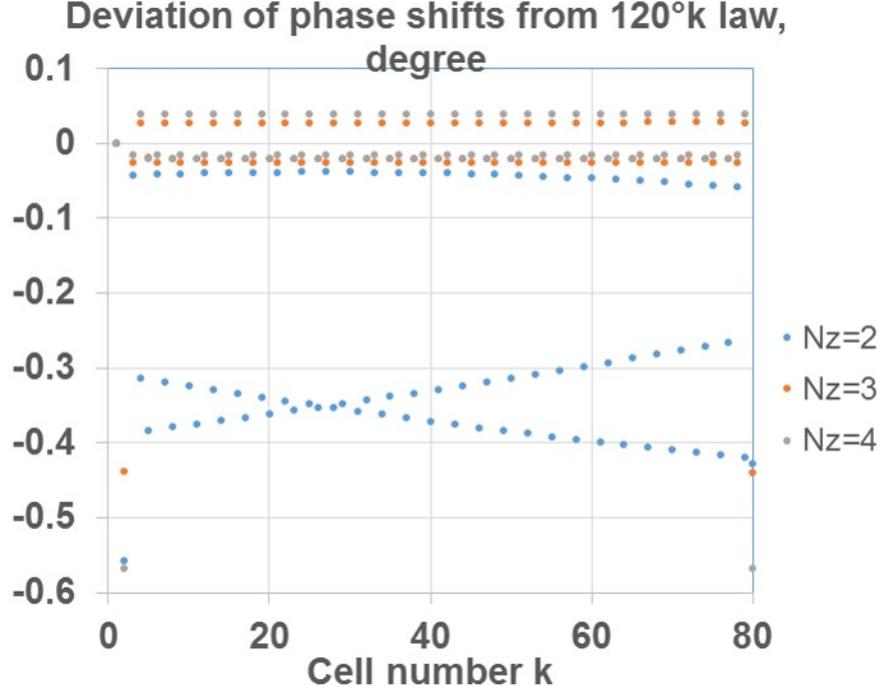

Figure 3: Deviation of phase shifts at centers of large waveguides from $120°k$ law for the case of homogeneous structure

It can be seen that with the appropriate choice of $N_Z$ and $N_R$ we get the correct results. Thus, the developed model can be used to study the properties of slow wave structures based on disk-loaded waveguides.

| | | | Table 1 Calculated dispersive characteristics | | | |
|---|---|---|---|---|---|---|
| | | | Phase shift $\varphi$ per cell –frequency f(MHz) | | | |
| | $N_Z$ | $N_R$ | $\varphi = 0$ | $\varphi = \pi/3$ | $\varphi = 2\pi/3$ | $\varphi = \pi$ |
| SUPERFISH | | | 2805.44 | 2822.33 | 2855.99 | 2872.77 |
| Bessel | 4 | 35 | 2805.34 | 2822.19 | 2855.83 | 2872.52 |
| Bessel | 4 | 70 | 2805.37 | 2822.26 | 2855.88 | 2872.60 |
| Legendre | 4 | 10 | 2805.42 | 2822.32 | 2855.94 | 2872.71 |
| Legendre | 4 | 25 | 2805.39 | 2822.27 | 2855.90 | 2872.65 |

# 5 Appendex 1

Solution of equation (2) is

$$E_{r,m}^{(q)} = B_{m,0}^{(q,1)} \exp\left(\gamma_m^{(q)}\tilde{z}\right) + B_{m,0}^{(q,2)} \exp\left(-\gamma_m^{(q)}\tilde{z}\right) + \frac{1}{\gamma_m^{(q)}} \int_0^{\tilde{z}} sh\left\{\gamma_m^{(q)}(\tilde{z} - \tilde{z}')\right\} F_m^{(q)}(\tilde{z}')d\tilde{z}', \qquad (42)$$

where $B_{m,0}^{(q,1)}$ and $B_{m,0}^{(q,2)}$ are constants. Integrals with $F_m^{(q)}$ can be simplified

$$\int_0^{\tilde{z}} sh\left\{\gamma_m^{(q)}(\tilde{z}-\tilde{z}')\right\} F_m^{(k)}(\tilde{z}')d\tilde{z}' = \frac{1}{i\,\omega\varepsilon_0\varepsilon}\frac{\lambda_m}{b_q} sh\left(\gamma_m^{(q)}\tilde{z}\right) I_{z,m}^{(q)}(0) - $$
$$-\gamma_m^{(q)2}\frac{1}{i\,\omega\varepsilon_0\varepsilon}\left[\int_0^{\tilde{z}} sh\left\{\gamma_m^{(q)}(\tilde{z}-\tilde{z}')\right\} I_{r,m}^{(q)}(\tilde{z}')d\tilde{z}' + \frac{\lambda_m}{b_q\gamma_m^{(q)}}\int_0^{\tilde{z}} ch\left\{\gamma_m^{(q)}(\tilde{z}-\tilde{z}')\right\} I_{z,m}^{(q)}(\tilde{z}')d\tilde{z}'\right] \quad (43)$$

$$\int_0^{\tilde{z}} ch\left\{\gamma_m^{(q)}(\tilde{z}-\tilde{z}')\right\} F_m^{(q)}(z_k+\tilde{z})d\tilde{z}' = -\frac{1}{i\,\omega\varepsilon_0\varepsilon}\frac{\lambda_m}{b_q} I_{z,m}^{(q)}(\tilde{z}) + \frac{1}{i\,\omega\varepsilon_0\varepsilon}\frac{\lambda_m}{b_q} I_{z,m}^{(q)}(0)\,ch\left(\gamma_m^{(q)}\tilde{z}\right) - $$
$$-\frac{\gamma_m^{(q)2}}{i\,\omega\varepsilon_0\varepsilon}\left[\int_0^{\tilde{z}} ch\left\{\gamma_m^{(q)}(\tilde{z}-\tilde{z}')\right\} I_{r,m}^{(q)}d\tilde{z}' + \frac{\lambda_m}{b_q\gamma_m^{(q)}}\int_0^{\tilde{z}} sh\left\{\gamma_m^{(q)}(\tilde{z}-\tilde{z}')\right\} I_{z,m}^{(q)}d\tilde{z}'\right] \quad (44)$$

Then (42) and (3) we can rewrite as [3]

$$E_{r,m}^{(q)} = B_{m,0}^{(q,1)}\exp\left(\gamma_m^{(k)}\tilde{z}\right) + B_{m,0}^{(q,2)}\exp\left(-\gamma_m^{(q)}\tilde{z}\right) + \frac{\gamma_m^{(q)}}{i\,\omega\varepsilon_0\varepsilon}\frac{\lambda_m}{b_q\gamma_m^{(q)2}} sh\left(\gamma_m^{(q)}\tilde{z}\right) I_{z,m}^{(q)}(0) - $$
$$\frac{\gamma_m^{(q)}}{i\,\omega\varepsilon_0\varepsilon}\left[\int_0^{\tilde{z}} sh\left\{\gamma_m^{(q)}(\tilde{z}-\tilde{z}')\right\} I_{r,m}^{(q)}(\tilde{z}')d\tilde{z}' + \frac{\lambda_m}{b_q\gamma_m^{(q)}}\int_0^{\tilde{z}} ch\left\{\gamma_m^{(q)}(\tilde{z}-\tilde{z}')\right\} I_{z,m}^{(q)}(\tilde{z}')d\tilde{z}'\right], \quad (45)$$

$$H_{\varphi,m}^{(q)} = +\frac{i\,\omega\varepsilon_0\varepsilon}{\gamma_m^{(q)}}\left\{B_{m,0}^{(q,1)}\exp\left(\gamma_m^{(q)}\tilde{z}\right) - B_{m,0}^{(q,2)}\exp\left(-\gamma_m^{(q)}\tilde{z}\right)\right\} + \frac{\lambda_m}{b_q\gamma_m^{(q)2}} I_{z,m}^{(q)}(0)\,ch\left(\gamma_m^{(q)}\tilde{z}\right) - $$
$$\left[\int_0^{\tilde{z}} ch\left\{\gamma_m^{(q)}(\tilde{z}-\tilde{z}')\right\} I_{r,m}^{(q)}d\tilde{z}' + \frac{\lambda_m}{b_q\gamma_m^{(q)}}\int_0^{\tilde{z}} sh\left\{\gamma_m^{(q)}(\tilde{z}-\tilde{z}')\right\} I_{z,m}^{(q)}d\tilde{z}'\right] \quad (46)$$

$$E_{z,m}^{(q)} = -\frac{\lambda_m}{b_q\gamma_m^{(q)}}\left\{B_{m,0}^{(q,1)}\exp\left(\gamma_m^{(q)}\tilde{z}\right) - B_{m,0}^{(q,2)}\exp\left(-\gamma_m^{(q)}\tilde{z}\right)\right\} + $$
$$\frac{1}{i\,\omega\varepsilon_0\varepsilon}\frac{\lambda_m}{b_q}\left\{\frac{b_q}{\lambda_m} I_{z,m}^{(q)}(\tilde{z}) - \frac{\lambda_m}{b_q\gamma_m^{(q)2}} I_{z,m}^{(q)}(0)\,ch\left(\gamma_m^{(q)}\tilde{z}\right)\right\} - $$
$$\frac{1}{i\,\omega\varepsilon_0\varepsilon}\frac{\lambda_m}{b_q}\left[\int_0^{\tilde{z}} ch\left\{\gamma_m^{(q)}(\tilde{z}-\tilde{z}')\right\} I_{r,m}^{(q)}d\tilde{z}' + \frac{\lambda_m}{b_q\gamma_m^{(q)}}\int_0^{\tilde{z}} sh\left\{\gamma_m^{(q)}(\tilde{z}-\tilde{z}')\right\} I_{z,m}^{(q)}d\tilde{z}'\right] \quad (47)$$

Consider the subdomains with $2 \leqslant k \leqslant (N_{REZ}-1)$ and $2 \leqslant k' \leqslant N_{REZ}$. From the conditions of the continuity of the electric fields at the interfaces between subdomains we get

$$\begin{cases} E_{r,m}^{(k')}(0) = \dfrac{2}{J_1^2(\lambda_m)}\sum_s C_s^{(k_2-1)} R_{m,s}^{\varphi,r}, \\ E_{r,m}^{(k')}(d_{k'}) = \dfrac{2}{J_1^2(\lambda_m)}\sum_s C_s^{(k_1)} R_{m,s}^{\varphi,r(k',k)}, \end{cases} \quad (48)$$

---
[3] Dimension of $I_{z,m}^{(q)}\left(I_{r,m}^{(q)}\right)$ is $A/m^2$

$$\begin{cases} E_{r,m}^{(k_1)}(0) = \dfrac{2b_{k'}^2}{b_k^2 J_1^2(\lambda_m)} \sum_s C_s^{(k_1)} R_{m,s}^{\varphi,r(k',k)} \\ E_{z,m}^{(k_1)}(d_k/2) = \dfrac{2}{J_1^2(\lambda_m)} \sum_s Q_s^{(k)} R_{m,s}^{\varphi,z} \end{cases} \qquad (49)$$

$$\begin{cases} E_{z,m}^{(k_2)}(0) = \dfrac{2}{J_1^2(\lambda_m)} \sum_s Q_s^{(k)} R_{m,s}^{\varphi,z}, \\ E_{r,m}^{(k_2)}(d_k/2) = \dfrac{2b_{k'+1}^2}{b_k^2 J_1^2(\lambda_m)} \sum_s C_s^{(k_2)} R_{m,s}^{\varphi,r(k'+1,k)} \end{cases} \qquad (50)$$

From these equations, one can obtain the field amplitudes in the subdomains

$$\begin{aligned}
B_{m,0}^{(k',1)} &= -\tilde{B}_{m,0}^{(k',1)} + \frac{1}{sh\left(\gamma_m^{(k')} d_{k'}\right) J_1^2(\lambda_m)} \sum_s C_s^{(k_1)} R_{m,s}^{\varphi,r(k',k')} - \\
\exp\left(-\gamma_m^{(k')} d_{k'}\right) & \frac{1}{sh\left(\gamma_m^{(k')} d_{k'}\right) J_1^2(\lambda_m)} \sum_s C_s^{(k_2-1)} R_{m,s}^{\varphi,r(k',k')} \\
B_{m,0}^{(k',2)} &= \tilde{B}_{m,0}^{(k',1)} - \frac{1}{sh\left(\gamma_m^{(k')} d_{k'}\right) J_1^2(\lambda_m)} \sum_s C_s^{(k_1)} R_{m,s}^{\varphi,r(k',k')} + \\
\exp\left(\gamma_m^{(k')} d_{k'}\right) & \frac{1}{sh\left(\gamma_m^{(k')} d_{k'}\right) J_1^2(\lambda_m)} \sum_s C_s^{(k_2-1)} R_{m,s}^{\varphi,r(k',k')}
\end{aligned} \qquad (51)$$

$$\begin{aligned}
B_{m,0}^{(k_1,1)} &= \tilde{B}_{m,0}^{(k_1,1)} - \frac{b_k \gamma_m^{(k)}}{\lambda_m J_1^2(\lambda_m) ch\left(\gamma_m^{(k)} d_k/2\right)} \sum_s Q_s^{(k)} R_{m,s}^{\varphi,z} + \\
\frac{b_{k'}^2 \exp\left(-\gamma_m^{(k)} d_k/2\right)}{b_k^2 J_1^2(\lambda_m) ch\left(\gamma_m^{(k)} d_k/2\right)} & \sum_s C_s^{(k_1)} R_{m,s}^{\varphi,r(k',k)} \\
B_{m,0}^{(k_1,2)} &= -\tilde{B}_{m,0}^{(k_1,1)} + \frac{b_k \gamma_m^{(k)}}{\lambda_m J_1^2(\lambda_m) ch\left(\gamma_m^{(k)} d_k/2\right)} \sum_s Q_s^{(k)} R_{m,s}^{\varphi,z} + \\
\frac{b_{k'}^2 \exp\left(\gamma_m^{(k)} d_k/2\right)}{b_k^2 J_1^2(\lambda_m) ch\left(\gamma_m^{(k)} d_k/2\right)} & \sum_s C_s^{(k_1)} R_{m,s}^{\varphi,r(k',k)}
\end{aligned} \qquad (52)$$

$$B_{m,0}^{(k_2,1)} = -\tilde{B}_{m,0}^{(k_2,1)} - \frac{\exp\left(-\gamma_m^{(k)}d_k/2\right)}{i\,\omega\varepsilon_0}\frac{b_k}{2\gamma_m^{(k)}\lambda_m ch\left(\gamma_m^{(k)}d_k/2\right)}\frac{\omega^2}{c^2}I_{z,m}^{(k_2)}(0) +$$

$$+\frac{b_{k'+1}^2}{b_k^2 ch\left(\gamma_m^{(k)}d_k/2\right)J_1^2(\lambda_m)}\sum_s C_s^{(k_2)} R_{m,s}^{\varphi,r(k'+1,k)} - \frac{\exp\left(-\gamma_m^{(k)}d_k/2\right)}{ch\left(\gamma_m^{(k)}d_k/2\right)J_1^2(\lambda_m)}\frac{b_k\gamma_m^{(k)}}{\lambda_m}\sum_s Q_s^{(k)} R_{m,s}^{\varphi,z}$$

$$B_{m,0}^{(k_2,2)} = -\tilde{B}_{m,0}^{(k_2,1)} + \frac{\exp\left(\gamma_m^{(k)}d_k/2\right)}{i\,\omega\varepsilon_0}\frac{b_k}{2\gamma_m^{(k)}\lambda_m ch\left(\gamma_m^{(k)}d_k/2\right)}\frac{\omega^2}{c^2}I_{z,m}^{(k_2)}(0) +$$

$$+\frac{b_{k'+1}^2}{b_k^2 ch\left(\gamma_m^{(k)}d_k/2\right)J_1^2(\lambda_m)}\sum_s C_s^{(k_2)} R_{m,s}^{\varphi,r(k'+1,k)} + \frac{\exp\left(\gamma_m^{(k)}d_k/2\right)}{ch\left(\gamma_m^{(k)}d_k/2\right)J_1^2(\lambda_m)}\frac{b_k\gamma_m^{(k)}}{\lambda_m}\sum_s Q_s^{(k)} R_{m,s}^{\varphi,z}$$

(53)

where

$$\tilde{B}_{m,0}^{(k',1)} = \frac{1}{i\,\omega\varepsilon_0\varepsilon}\frac{\lambda_m}{2\gamma_m^{(k')}b_{k'}}I_{z,m}^{(k')}(0) -$$

$$\frac{1}{i\,\omega\varepsilon_0\varepsilon}\frac{\gamma_m^{(k')}}{2sh\left(\gamma_m^{(k')}d_{k'}\right)}\left\{\begin{array}{l}\int_0^{d_{k'}} sh\left\{\gamma_m^{(k')}(d_{k'}-\tilde{z}')\right\} I_{r,m}^{(k')}(\tilde{z}')d\tilde{z}' \\ +\frac{\lambda_m}{b_{k'}\gamma_m^{(k')}}\int_0^{d_{k'}} ch\left\{\gamma_m^{(k')}(d_{k'}-\tilde{z}')\right\} I_{z,m}^{(k')}(\tilde{z}')d\tilde{z}'\end{array}\right\}$$

(54)

$$2\tilde{B}_{m,0}^{(k_1,1)}ch\left(\gamma_m^{(k)}d_k/2\right) = \frac{1}{i\,\omega\varepsilon_0\varepsilon}\frac{b_k\gamma_m^{(k)}}{\lambda_m}I_{z,m}^{(k_1)}(d_k/2) - \frac{1}{i\,\omega\varepsilon_0\varepsilon}\frac{\lambda_m}{\gamma_m^{(k)}b_k}I_{z,m}^{(k_1)}(0)ch\left(\gamma_m^{(k)}d_k/2\right) +$$

$$+\frac{\gamma_m^{(k)}}{i\,\omega\varepsilon_0\varepsilon}\left\{\int_0^{d_k/2} ch\left\{\gamma_m^{(k)}(d_k/2-\tilde{z}')\right\} I_{r,m}^{(k_1)}d\tilde{z}' + \frac{\lambda_m}{b_k\gamma_m^{(k)}}\int_0^{d_k/2} sh\left\{\gamma_m^{(k)}(d_k/2-\tilde{z}')\right\} I_{z,m}^{(k_1)}d\tilde{z}'\right\}$$

(55)

$$2\tilde{B}_{m,0}^{(k_2,1)}ch\left(\gamma_m^{(k)}d_k/2\right) = \frac{1}{i\,\omega\varepsilon_0\varepsilon}\frac{\lambda_m}{b_k\gamma_m^{(k)}}sh\left(\gamma_m^{(k)}d_k/2\right) I_{z,m}^{(k_2)}(0) -$$

$$\frac{\gamma_m^{(k)}}{i\,\omega\varepsilon_0\varepsilon}\left\{\begin{array}{l}\int_0^{d_k/2} sh\left\{\gamma_m^{(k)}(d_k/2-\tilde{z}')\right\} I_{r,m}^{(k_2)}(\tilde{z}')d\tilde{z}' + \\ +\frac{\lambda_m}{b_k\gamma_m^{(k)}}\int_0^{d_k/2} ch\left\{\gamma_m^{(k)}(d_k/2-\tilde{z}')\right\} I_{z,m}^{(k_2)}(\tilde{z}')d\tilde{z}'\end{array}\right\}$$

(56)

If we substitute (51)-(53) into another continuity conditions

$$\sum_m H_{\varphi,m}^{(k')}(d_{k'}) R_{s',m}^{\psi(k',k')} = \sum_m H_{\varphi,m}^{(k_1)}(0) R_{s',m}^{\psi(k',k)},$$

$$\sum_m H_{\varphi,m}^{(k_2)}(d_k/2) R_{s',m}^{\psi(k'+1,k)} = \sum_m H_{\varphi,m}^{(k'+1)}(0) R_{s',m}^{\psi(k'+1,k'+1)},$$

(57)

$$E_{r,m}^{(k_1)}(d_k/2) = E_{r,m}^{(k_2)}(0), \tag{58}$$

we get the seeking equations for vectors $C^{(k_1)}$, $C^{(k_2)}$ and $Q^{(k)}$

$$-\sum_s T_{s',s}^{\prime(1,k')} C_s^{(k_2-1)} + \sum_s \left( T_{s',s}^{\prime(2,k')} b_{k'} + T_{s',s}^{(2,k',k)} \right) C_s^{(k_1)} + \sum_s T_{s',s}^{(1,k',k)} Q_s^{(k)} = Z_{s'}^{-(k)},$$

$$-\sum_s \left( T_{s',s}^{\prime(2,k'+1)} + T_{s',s}^{(2,k'+1,k)} \right) C_s^{(k_2)} + \sum_s T_{s',s}^{\prime(1,k'+1)} C_s^{(k_1+1)} + \sum_s T_{s',s}^{(1,k',k)} Q_s^{(k)} = Z_{s'}^{+(k)}, \tag{59}$$

$$\sum_s T_{m,s}^{r(k'+1,k)} C_s^{(k_2)} - \sum_s T_{m,s}^{r(k',k)} C_s^{(k_1)} + \sum_s T_{m,s}^{z(k)} Q_s^{(k)} = Z_m^{(k)},$$

where

$$T_{s',s}^{\prime(1,k')} = \frac{b_{k'}}{b_*} \sum_m \frac{2}{\gamma_m^{(k')} b_{k'} sh\left(\gamma_m^{(k')} d_{k'}\right) J_1^2(\lambda_m)} R_{s',m}^{\psi(k',k')} R_{m,s}^{\varphi,r(k',k')}$$

$$T_{s',s}^{\prime(2,k')} = \frac{b_{k'}}{b_*} \sum_m \frac{2 ch\left(\gamma_m^{(k')} d_{k'}\right)}{\gamma_m^{(k')} b_{k'} sh\left(\gamma_m^{(k')} d_{k'}\right) J_1^2(\lambda_m)} R_{s',m}^{\psi(k',k')} R_{m,s}^{\varphi,r(k',k')}$$

$$T_{s',s}^{(2,k',k)} = \frac{b_{k'}}{b_*} \frac{b_{k'}}{b_k} \sum_m \frac{2 sh\left(\gamma_m^{(k)} d_k/2\right)}{\gamma_m^{(k)} b_k ch\left(\gamma_m^{(k)} d_k/2\right) J_1^2(\lambda_m)} R_{s',m}^{\psi(k',k)} R_{m,s}^{\varphi,r(k',k)} \tag{60}$$

$$T_{s',s}^{(1,k',k)} = \frac{b_k}{b_*} \sum_m \frac{2}{\lambda_m} \frac{1}{ch\left(\gamma_m^{(k)} d_k/2\right) J_1^2(\lambda_m)} R_{s',m}^{\psi(k',k)} R_{m,s}^{\varphi,z}$$

$$T_{s',s}^{r(k',k)} = \frac{\lambda_{s'}}{2 b_k \gamma_{s'}^{(k)} sh\left(\gamma_{s'}^{(k)} d_k/2\right)} \frac{b_{k'}^2}{b_k^2} R_{s',s}^{\varphi,r(k',k)},$$

$$T_{s',s}^{z(k)} = R_{s',s}^{\varphi,z},$$

$$R_{m,s}^{\varphi,r(k',k)} = \int_0^1 \varphi_s^{(r)}(x) J_1(b_{k'} \lambda_m x/b_k) x dx,$$

$$R_{m,s}^{\varphi,z} = \int_0^1 \varphi_s^{(z)}(x) J_0(\lambda_m x) x dx, \tag{61}$$

$$R_{s',m}^{\psi(k',k)} = \int_0^1 \psi_{s'}(x) J(b_{k'} \lambda_m x/b_k) x dx,$$

$$i\,\omega\varepsilon_0\varepsilon b_* Z_s^{(k)} = \frac{b_*\lambda_s J_1^2(\lambda_s)}{4 b_k sh\left(\gamma_s^{(k)} d_k/2\right)} \left[ \begin{cases} \int\limits_0^{d_k/2} sh\left\{\gamma_s^{(k)} \tilde z'\right\} I_{r,s}^{(k_1)} d\tilde z' - \\ \frac{\lambda_s}{b_k \gamma_s^{(k)}} \int\limits_0^{d_k/2} ch\left\{\gamma_s^{(k)} \tilde z'\right\} I_{z,s}^{(k_1)} d\tilde z' \end{cases} - \\ \begin{cases} \int\limits_0^{d_k/2} sh\left\{\gamma_s^{(k)}(d_k/2 - \tilde z')\right\} I_{r,s}^{(k_2)}(\tilde z') d\tilde z' - \\ \frac{\lambda_s}{b_k \gamma_s^{(k)}} \int\limits_0^{d_k/2} ch\left\{\gamma_s^{(k)}(d_k/2 - \tilde z')\right\} I_{z,s}^{(k_2)}(\tilde z') d\tilde z' \end{cases} \right] \tag{62}$$

$$i\,\omega\varepsilon_0\varepsilon b_* Z_{s'}^{-(k)} = \sum_m \frac{R_{s',m}^{\psi(k',k)}}{ch\left(\gamma_m^{(k)} d_k/2\right)} \left\{ \begin{array}{l} \int\limits_0^{d_k/2} ch\left\{\gamma_m^{(k)}(d_k/2 - \tilde z')\right\} I_{r,m}^{(k_1)} d\tilde z' + \\ \frac{\lambda_m}{\gamma_m^{(k)} b_k} \int\limits_0^{d_k/2} sh\left\{\gamma_m^{(k)}(d_k/2 - \tilde z')\right\} I_{z,m}^{(k_1)} d\tilde z' \end{array} \right\} +$$
$$+ \sum_m \frac{b_k I_{z,m}^{(k_1)}(d_k/2) R_{s',m}^{\psi(k',k)}}{\lambda_m ch\left(\gamma_m^{(k)} d_k/2\right)} + \sum_m \frac{R_{s',m}^{\psi(k',k')}}{sh\left(\gamma_m^{(k')} d_{k'}\right)} \left\{ \begin{array}{l} \int\limits_0^{d_{k'}} sh\left(\gamma_m^{(k')} \tilde z'\right) I_{r,m}^{(k')}(\tilde z') d\tilde z' - \\ \frac{\lambda_m}{\gamma_m^{(k')} b_{k'}} \int\limits_0^{d_{k'}} ch\left(\gamma_m^{(k')} \tilde z'\right) I_{z,m}^{(k')}(\tilde z') d\tilde z' \end{array} \right\} \tag{63}$$

$$i\,\omega\varepsilon_0\varepsilon b_* Z_{s'}^{+(k)} = \sum_m \frac{b_k I_{z,m}^{(k_2)}(0) R_{s',m}^{\psi((k+1)',k)}}{\lambda_m ch\left(\gamma_m^{(k)} d_k/2\right)} -$$
$$\sum_m \frac{R_{s',m}^{\psi((k+1)',(k+1)')}}{sh\left(\gamma_m^{((k+1)')} d_{(k+1)'}\right)} \left\{ \begin{array}{l} \int\limits_0^{d_{(k+1)'}} sh\left\{\gamma_m^{((k+1)')}(d_{(k+1)'} - \tilde z')\right\} I_{r,m}^{((k+1)')}(\tilde z') d\tilde z' + \\ \frac{\lambda_m}{b_{(k+1)'}\gamma_m^{((k+1)')}} \int\limits_0^{d_{(k+1)'}} ch\left\{\gamma_m^{((k+1)')}(d_{(k+1)'} - \tilde z')\right\} I_{z,m}^{((k+1)')}(\tilde z') d\tilde z' \end{array} \right\} - \tag{64}$$
$$\sum_m \frac{R_{s',m}^{\psi((k+1)',k)}}{ch\left(\gamma_m^{(k)} d_k/2\right)} \left\{ \begin{array}{l} \int\limits_0^{d_k/2} ch\left(\gamma_m^{(k)} \tilde z'\right) I_{r,m}^{(k_2)}(\tilde z') d\tilde z' + \\ -\frac{\lambda_m}{b_k \gamma_m^{(k)}} \int\limits_0^{d_k/2} sh\left(\gamma_m^{(k)} \tilde z'\right) I_{z,m}^{(k_2)}(\tilde z') d\tilde z' \end{array} \right\}$$

Consider the case when the dimensions of two waveguides are chosen such that only the dominant mode propagates, and the higher-order modes are all evanescent We will suppose that there is an incident wave that travels from $z = -\infty$ with amplitude $G_1^{(1)} = 1$ ($G_s^{(1)} = 0$, $s \geqslant 2$). The boundary condition for electric field at the junction of the first semi-infinite circular waveguide gives the following equality

$$\sum_s \left( G_s^{(1)} \mathcal{E}_{s,r}^{(1)} + G_{-s}^{(1)} \mathcal{E}_{-s,r}^{(1)} \right) = \begin{cases} \sum_{s'} C_{s'}^{(L)} \varphi_{s'} \left( \dfrac{r}{b_{1'}} \right), & 0 \leqslant r \leqslant b_{1'}, \\ 0, & b_{1'} \leqslant r < b_{w_1}. \end{cases} \quad (65)$$

The right hand side of the equation (65) we expand in terms of a set of orthogonal and complete functions $J_1 \left( \frac{\lambda_s}{b_{w_1}} r \right)$. Equating the expansion coefficients gives

$$\begin{aligned}
G_{-1}^{(1)} &= 1 + 2 \frac{b_{1'}^2 \lambda_1}{J_1^2(\lambda_1) b_{w_1}^2 \gamma_1^{(w_1)} b_{w_1}} \sum_{s'} R_{1,s'}^{w,L} C_{s'}^{(L)}, \\
G_{-s}^{(1)} &= 2 \frac{b_{1'}^2 \lambda_s}{J_1^2(\lambda_s) b_{w_1}^2 \gamma_s^{(w_1)} b_{w_1}} \sum_{s'} R_{s,s'}^{w,L} C_{s'}^{(L)}, \quad s = 2, 3, ...
\end{aligned} \quad (66)$$

Using the same procedure at the junction of the second semi-infinite circular waveguide $E_r = \sum_{s'} C_{s'}^{(R)} \varphi_{s'} \left( r/b_{(N_{REZ}+1)'} \right)$, we get

$$G_s^{(2)} = -2 \frac{b_{(N_R+1)'}^2 \lambda_s}{J_1^2(\lambda_s) b_{w_2}^2 \gamma_s^{(w_2)} b_{w_2}} \sum_{s'} R_{s,s'}^{w,R} C_{s'}^{(R)}, \quad s = 1, 2, ... \quad (67)$$

where

$$\begin{aligned}
R_{m,s}^{w,L} &= \int_0^1 \varphi_s^{(r)}(x) J_1 \left( b_{1'} \lambda_m x / b_{w_1} \right) x dx, \\
R_{m,s}^{w,R} &= \int_0^1 \varphi_s^{(r)}(x) J_1 \left( b_{(N_R+1)'} \lambda_m x / b_{w_2} \right) x dx.
\end{aligned} \quad (68)$$

The reflection and transmission coefficients are given by $R = G_{-1}^{(1)}$ and $T = G_1^{(2)}$.

The conditions for the continuity of magnetic fields

$$\begin{aligned}
\sum_s \left( G_s^{(1)} \mathcal{H}_{\varphi,s}^{(w,1)} + G_{-s}^{(1)} \mathcal{H}_{-s,\varphi}^{(w,1)} \right) &= \sum_m H_{\varphi,m}^{(1')}(0) J_1 \left( \dfrac{\lambda_m}{b_{1'}} r \right), \quad 0 < r < b_{1'},\; \tilde{z} = 0, \\
\sum_s G_s^{(2)} \mathcal{H}_{\varphi,s}^{(w,2)} &= \\
&\sum_m H_{\varphi,m}^{((N_R+1)')} \left( d_{(N_{REZ}+1)'} \right) J_1 \left( \dfrac{\lambda_m}{b_{(N_{REZ}+1)'}} r \right), \quad 0 < r < b_{(N_R+1)'},\; \tilde{z} = d_{(N_{REZ}+1)'}
\end{aligned} \quad (69)$$

give

$$-i\omega\varepsilon_0 \sum_s \frac{b_{w,1}}{\lambda_s} R_{s',s}^{\psi(w,1)} \left( G_s^{(1)} + G_{-s}^{(1)} \right) = \sum_s R_{s',s}^{\psi(1,1)} H_{\varphi,s}^{(1')}(0), \quad (70)$$

$$-i\omega\varepsilon_0 \sum_s \frac{b_{w,2}}{\lambda_s} R_{s',s}^{\psi(w,2)} G_s^{(2)} = \sum_s R_{s',s}^{\psi((N_{REZ}+1)',2)} H_{\varphi,m}^{((N_R+1)')} \left( d_{(N_{REZ}+1)'} \right), \quad (71)$$

where

$$R_{m',m}^{\psi(w,1)} = \int_0^1 \psi_{m'}(x) J_1\left(\frac{b_{1'}\lambda_m}{b_{w,1}}x\right) xdx,$$

$$R_{m',m}^{\psi(w,2)} = \int_0^1 \psi_{m'}(x) J_1\left(\frac{b_{(N_R+1)'}\lambda_m}{b_{w,2}}x\right) xdx. \tag{72}$$

Substituting (66),(67) into (70),(71) gives

$$-\varepsilon \sum_s T_{s',s}^{\prime(1,1')} C_s^{(1_1)} + \varepsilon \sum_s T_{s',s}^{\prime(2,1')} C_s^{(L)} - \sum_s T_{s',s}^{(L)} C_s^{(L)} = R_{s'}^{(L)} + Z_{s'}^L \tag{73}$$

$$\sum_s T_{s',s}^{(R)} C_s^{(R)} - \varepsilon \sum_s T_{s',s}^{\prime(2,(N_R+1)')} C_s^{(R)} + \varepsilon \sum_s T_{s',s}^{\prime(1,(N_R+1)')} C_s^{((N_R)_2)} = Z_{s'}^{(R)}, \tag{74}$$

where

$$i\omega\varepsilon_0 b_{(N_R+1)'} Z_{s'}^{(R)} =$$

$$-\sum_m \frac{R_{s',m}^{\psi((N_R+1)',(N_R+1)')}}{sh\left(\gamma_m^{((N_R+1)')} d_{(N_R+1)'}\right)} \left\{ \begin{array}{l} \int_0^{d_{k'}} sh\left\{\gamma_m^{((N_R+1)')} \tilde{z}'\right\} I_{r,m}^{((N_R+1)')} d\tilde{z}' - \\ \\ \frac{\lambda_m}{b_{(N_R+1)'}\gamma_m^{((N_R+1)')}} \int_0^{d_{k'}} ch\left\{\gamma_m^{((N_R+1)')} \tilde{z}'\right\} I_{z,m}^{((N_R+1)')} d\tilde{z}' \end{array} \right\}, \tag{75}$$

$$i\omega\varepsilon_0 b_* Z_{s'}^L = \sum_m \frac{R_{s',m}^{\psi(1',1')}}{sh\left(\gamma_m^{(1')} d_{1'}\right)} \times \left\{ \begin{array}{l} \int_0^{d_{k'}} sh\left\{\gamma_m^{(1')}(d_{1'} - \tilde{z}')\right\} I_{r,m}^{(1')}(\tilde{z}') d\tilde{z}' \\ \\ + \frac{\lambda_m}{b_{1'}\gamma_m^{(1')}} \int_0^{d_{k'}} ch\left\{\gamma_m^{(1')}(d_{1'} - \tilde{z}')\right\} I_{z,m}^{(1')}(\tilde{z}') d\tilde{z}' \end{array} \right\}, \tag{76}$$

$$T_{s',s}^{(L)} = 2\frac{b_{w,1}}{b_*}\frac{b_{1'}^2}{b_{w_1}^2}\sum_m \frac{1}{J_1^2(\lambda_m)\gamma_m^{(w_1)}b_{w_1}} R_{m,s}^{w,L} R_{s',m}^{\psi(w,1)},$$

$$R_{s'}^{(L)} = 2\frac{b_{w,1}}{b_*\lambda_1} R_{s',1}^{\psi(w,1)}. \tag{77}$$

# 6 Appendex 2

In this Appendix, we give a procedure for transforming a system of vector equations

$$-\varepsilon T^{\prime(1,1')} C^{(1_1)} + \varepsilon T^{\prime(2,1')} C^{(L)} - T^{(L)} C^{(L)} = R^{(L)} + Z^{(L)}, \tag{78}$$

$$\left(T^{\prime(2,k')} + T^{(2,k',k)}\right) C^{(k_1)} = T^{\prime(1,k')} C^{((k-1)_2)} - T^{(1,k',k)} Q^{(k)} + Z^{-(k)}, k = 1, 2, ..., N_{REZ}, \tag{79}$$

$$\left(T^{\prime(2,(k+1)')} + T^{(2,(k+1)',k)}\right) C^{(k_2)} =$$
$$T^{\prime(1,(k+1)')} C^{((k+1)_1)} + T^{(1,(k+1)',k)} Q^{(k)} - Z^{+(k)}, k = 1, 2, ..., N_{REZ}, \tag{80}$$

$$T^{r((k+1)',k)}C^{(k_2)} - T^{r(k',k)}C^{(k_1)} + T^{z(k)}Q^{(k)} = Z^{(k)}, k = 1, 2, ..., N_{REZ}, \tag{81}$$

$$T^{(R)}C^{(R)} - \varepsilon T'^{(2,(N_{REZ}+1)')}C^{(R)} + \varepsilon T'^{(1,(N_{REZ}+1)')}C^{((N_{REZ})_2)} = Z^{(R)}, \tag{82}$$

into a system including only vectors $Q^{(k)}$.

Equation (79) can be rewritten as

$$\left(T'^{(2,1')} + T^{(2,1',1)}\right)C^{(1_1)} = T'^{(1,1')}C^{((0)_2)} - T^{(1,1',1)}Q^{(1)} + Z^{-(1)}, \tag{83}$$

$$\left(T'^{(2,k')} + T^{(2,k',k)}\right)C^{(k_1)} = T'^{(1,k')}C^{((k-1)_2)} - T^{(1,k',k)}Q^{(k)} + Z^{-(k)}, k = 2, ..., N_{REZ}. \tag{84}$$

Shifting the indexes in (84), we get

$$\left(T'^{(2,k'+1)} + T^{(2,k'+1,k+1)}\right)C^{((k+1)_1)} = \\ T'^{(1,k'+1)}C^{((k)_2)} - T^{(1,k'+1,k+1)}Q^{(k+1)} + Z^{-(k+1)}, k = 1, ..., N_{REZ} - 1. \tag{85}$$

Equation (80) is also divided into two equations

$$\left(T'^{(2,(k+1)')} + T^{(2,(k+1)',k)}\right)C^{(k_2)} = \\ T'^{(1,(k+1)')}C^{((k+1)_1)} + T^{(1,(k+1)',k)}Q^{(k)} - Z^{+(k)}, k = 1, ..., N_{REZ} - 1, \tag{86}$$

$$\left(T'^{(2,(N_{REZ}+1)')} + T^{(2,(N_{REZ}+1)',N_{REZ})}\right)C^{((N_{REZ})_2)} = \\ T'^{(1,(N_{REZ}+1)')}C^{((N_{REZ}+1)_1)} + T^{(1,(N_{REZ}+1)',N_{REZ})}Q^{(N_{REZ})} - Z^{+(N_{REZ})}. \tag{87}$$

From (85)) and (86) we get

$$C^{(k_2)} = T^{Ck_2Q(k)}Q^{(k)} + T^{Ck_2Q(k+1)}Q^{(k+1)} + Z^{Ck_2}, k = 1, ..., N_{REZ} - 1. \tag{88}$$

where

$$T^{Ck_2Q(k)} = W^{(1,k)}T^{(1,(k+1)',k)}$$

$$T^{Ck_2Q(k+1)} = -W^{(1,k)}T'^{(1,(k+1)')}\left(T'^{(2,(k+1)')} + T^{(2,(k+1)',k+1)}\right)^{-1}T^{(1,(k+1)',k+1)}$$

$$Z^{Ck_2} = W^{(1,k)}\left\{T'^{(1,(k+1)')}\left(T'^{(2,(k+1)')} + T^{(2,(k+1)',k+1)}\right)^{-1}Z^{-(k+1)} - Z^{+(k)}\right\}$$

$$W^{(1,k)} = \left\{\begin{array}{l}\left(T'^{(2,(k+1)')} + T^{(2,(k+1)',k)}\right) - \\ T'^{(1,(k+1)')}\left(T'^{(2,(k+1)')} + T^{(2,(k+1)',k+1)}\right)^{-1}T'^{(1,(k+1)')}\end{array}\right\}^{-1} \tag{89}$$

Shifting indexes in (86), we get

$$\left(T'^{(2,k')} + T^{(2,k',k)}\right)C^{((k-1)_2)} = T'^{(1,k')}C^{((k)_1)} + T^{(1,k',k-1)}Q^{(k-1)} - Z^{+(k-1)}, k = 2, ..., N_{REZ}. \tag{90}$$

From (90) and (84) we get

$$C^{(k_1)} = T^{Ck_1Q(k-1)}Q^{(k-1)} + T^{Ck_1Q(k)}Q^{(k)} + Z^{Ck_1}, k = 2, ..., N_{REZ}, \tag{91}$$

where

$$T^{Ck_1Q(k-1)} = W^{(2,k)}T'^{(1,(k)')}\left(T'^{(2,(k)')} + T^{(2,(k)',k-1)}\right)^{-1}T^{(1,(k)',k-1)}$$

$$T^{Ck_1Q(k)} = -W^{(2,k)}T^{(1,k',k)}$$

$$Z^{Ck_1} = W^{(2,k)}\left\{-T'^{(1,(k)')}\left(T'^{(2,(k)')} + T^{(2,(k)',k-1)}\right)^{-1}Z^{(k-1)} + Z^{-(k)}\right\} \quad (92)$$

$$W^{(2,k)} = \left\{\begin{array}{l}\left(T'^{(2,k')} + T^{(2,k',k)}\right) - \\ T'^{(1,(k)')}\left(T'^{(2,(k)')} + T^{(2,(k)',k-1)}\right)^{-1}T'^{(1,(k)')}\end{array}\right\}^{-1}$$

The equation (81) we rewrite as

$$T^{z(1)}Q^{(1)} = T^{r(1',1)}C^{(1_1)} - T^{r(2',1)}C^{(1_2)} + Z^{(1)}, \quad (93)$$

$$T^{z(k)}Q^{(k)} = T^{r(k',k)}C^{(k_1)} - T^{r((k+1)',k)}C^{(k_2)} + Z^{(k)}, \quad k = 2, ..., N_R - 1, \quad (94)$$

$$T^{z(N_{REZ})}Q^{(N_{REZ})} = T^{r((N_{REZ})',N_{REZ})}C^{((N_{REZ})_1)} - T^{r((N_{REZ}+1)',N_{REZ})}C^{((N_{REZ})_2)} + Z^{(N_{REZ})}. \quad (95)$$

Substituting (88) and (91) into (94), we get

$$T^{(k)}Q^{(k)} = T^{+(k)}Q^{(k+1)} + T^{-(k)}Q^{(k-1)} + Z^{Q(k)}, \quad k = 2, ..., N_{REZ} - 1, \quad (96)$$

where

$$\begin{aligned}T^{(k)} &= \left(T^{z(k)} - T^{r(k',k)}T^{Ck_1Q(k)} + T^{r((k+1)',k)}T^{Ck_2Q(k)}\right), \\ T^{-(k)} &= T^{r(k',k)}T^{Ck_1Q(k-1)}. \\ T^{+(k)} &= -T^{r((k+1)',k)}T^{Ck_2Q(k+1)}, \\ Z^{Q(k)} &= T^{r(k',k)}Z^{Ck_1} - T^{r((k+1)',k)}Z^{Ck_2} + Z^{(k)}.\end{aligned} \quad (97)$$

The difference equation (96) includes $(N_{REZ} - 2)$ relationships between $N_{REZ}$ unknown vectors. Additional relationships can be obtained from (78),(83),(93)

$$-\varepsilon T'^{(1,1')}C^{(1_1)} + \varepsilon T'^{(2,1')}C^{(L)} - T^{(L)}C^{(L)} = R^{(L)} + Z^{(L)}, \quad (98)$$

$$\left(T'^{(2,1')} + T^{(2,1',1)}\right)C^{(1_1)} = T'^{(1,1')}C^{(0_2)} - T^{(1,1',1)}Q^{(1)} + Z^{-(1)}, \quad (99)$$

$$T^{z(1)}Q^{(1)} = T^{r(1',1)}C^{(1_1)} - T^{r(2',1)}C^{(1_2)} + Z^{(1)}, \quad (100)$$

and (82),(87),(95)

$$T^{(R)}C^{(R)} - \varepsilon T'^{(2,(N_{REZ}+1)')}C^{(R)} + \varepsilon T'^{(1,(N_{REZ}+1)')}C^{((N_{REZ})_2)} = Z^{(R)}, \quad (101)$$

$$\left(T'^{(2,(N_{REZ}+1)')} + T^{(2,(N_{REZ}+1)',k)}\right)C^{((N_{REZ})_2)} = \\ T'^{(1,(N_{REZ}+1)')}C^{((N_{REZ}+1)_1)} + T^{(1,(N_{REZ}+1)',N_{REZ})}Q^{(N_{REZ})} - Z^{+(N_{REZ})}, \quad (102)$$

$$T^{z(N_{REZ})}Q^{(N_{REZ})} = T^{r((N_{REZ})',N_{REZ})}C^{((N_{REZ})_1)} - T^{r((N_{REZ}+1)',N_{REZ})}C^{((N_{REZ})_2)} + Z^{(N_{REZ})}. \quad (103)$$

We remind that $C^{(L)} = C^{(0_2)}$ and $C^{(R)} = C^{((N_{REZ}+1)_1)}$.

From the first two equations (98) and (99) we can find $C^{(L)}$ and $C^{(1_1)}$ (vectors of field expansion coefficients at the first diaphragm)

$$C^{(1_1)} = T^{C1_1Q(1)}Q^{(1)} + T^{C1_1RL}R^{(L)} + Z^{C1_1}, \quad (104)$$

$$C^{(L)} = T^{CLQ(1)}Q^{(1)} + T^{CLRL}R^{(L)} + Z^{CL}, \quad (105)$$

where

$$T^{C1_1Q(1)} = -\left\{\left(T'^{(2,1')} + T^{(2,1',1)}\right) - T'^{(1,1')}\left(\varepsilon T'^{(2,1')} - T^{(L)}\right)^{-1}\varepsilon T'^{(1,1')}\right\}^{-1}T^{(1,1',1)},$$

$$T^{C1_1RL} = \left\{\left(T'^{(2,1')} + T^{(2,1',1)}\right) - T'^{(1,1')}\left(\varepsilon T'^{(2,1')} - T^{(L)}\right)^{-1}\varepsilon T'^{(1,1')}\right\}^{-1} \times \quad (106)$$

$$T'^{(1,1')}\left(\varepsilon T'^{(2,1')} - T^{(L)}\right)^{-1},$$

$$Z^{C1_1} = T'^{(1,1')}\left(\varepsilon T'^{(2,1')} - T^{(L)}\right)^{-1}Z^{(L)} + Z^{-(1)},$$

$$T^{CLQ(1)} = \left(\varepsilon T'^{(2,1')} - T^{(L)}\right)^{-1}\varepsilon T'^{(1,1')}T^{C1_1Q(1)}$$

$$T^{CLRL} = \left\{\left(\varepsilon T'^{(2,1')} - T^{(L)}\right)^{-1}\varepsilon T'^{(1,1')}T^{C1_1RL} + \left(\varepsilon T'^{(2,1')} - T^{(L)}\right)^{-1}\right\} \quad (107)$$

$$Z^{CL} = \left(\varepsilon T'^{(2,1')} - T^{(L)}\right)^{-1}Z^{(L)} + \left(\varepsilon T'^{(2,1')} - T^{(L)}\right)^{-1}\varepsilon T'^{(1,1')}Z^{1_1}.$$

The third equation (100) have the unknown $C^{(1_2)}$; which we can express through $Q^{(1)}, Q^{(2)}$ by using the equation (88) for $k = 1$

$$C^{(1_2)} = T^{C1_2Q(1)}Q^{(1)} + T^{C1_2Q(2)}Q^{(2)} + Z^{C1_2}. \quad (108)$$

Substituting of (108) and (104) into (100) gives

$$T^{Q(1)}Q^{(1)} + T^{Q(2)}Q^{(2)} = Z^{Q(1)}, \quad (109)$$

where

$$T^{Q(1)} = \left(T^{z(1)} + T^{r(2',1)}T^{C1_2Q(1)} - T^{r(1',1)}T^{C1_1Q(1)}\right),$$

$$T^{Q(2)} = T^{r(2',1)}T^{C1_2Q(2)}, \quad (110)$$

$$Z^{Q(1)} = T^{r(1',1)}T^{C1_1RL}R^{(L)} + T^{r(1',1)}Z^{C1_1} + Z^{(1)} - T^{r(2',1)}Z^{1_2}.$$

Similarly, from the equations (101) and (102) we can find $C^{(R)}, C^{((N_{REZ})_2)}$ (vectors of field expansion coefficients at the last diaphragm)

$$C^{((N_{REZ})_2)} = T^{C(N_{REZ})_2Q(N_{REZ})}Q^{(N_{REZ})} + Z^{C(N_{REZ})_2}, \quad (111)$$

$$C^{(R)} = T^{CRQ(N_{REZ})}Q^{(N_{REZ})} + Z^{CR}, \quad (112)$$

where

$$T^{C(N_{REZ})_2Q(N_{REZ})} =$$

$$\begin{cases} \left(T'^{(2,(N_{REZ}+1)')} + T^{(2,(N_{REZ}+1)',N_{REZ})}\right) + \\ T'^{(1,(N_{REZ}+1)')}\left(T^{(R)} - \varepsilon T'^{(2,(N_R+1)')}\right)^{-1}\varepsilon T'^{(1,(N_R+1)')} \end{cases}^{-1} T^{(1,(N_{REZ}+1)',k)}, \quad (113)$$

$$Z^{C(N_{REZ})_2} = T'^{(1,(N_{REZ}+1)')}\left(T^{(R)} - \varepsilon T'^{(2,(N_{REZ}+1)')}\right)^{-1}Z^{(R)} - \tilde{Z}^{+(N_{REZ})},$$

$$T^{CRQ(N_{REZ})} = -\left(T^{(R)} - \varepsilon T'^{(2,(N_{REZ}+1)')}\right)^{-1} \varepsilon T'^{(1,(N_{REZ}+1)')} T^{C(N_{REZ})_2 Q(N_{REZ})}$$

$$Z^{CR} = -\left(T^{(R)} - \varepsilon T'^{(2,(N_{REZ}+1)')}\right)^{-1} \varepsilon T'^{(1,(N_{REZ}+1)')} Z^{C(N_{REZ})_2} + \left(T^{(R)} - \varepsilon T'^{(2,(N_{REZ}+1)')}\right)^{-1} Z^{(R)} \quad (114)$$

The third equation (103) have the unknown $C^{((N_{REZ})_1)}$, which we can express through $Q^{(N_{REZ}-1)}$, $Q^{(N_{REZ})}$ by using the equation (91) for $k = N_{REZ}$

$$C^{((N_{REZ})_1)} = T^{C(N_{REZ})_1 Q(N_{REZ}-1)} Q^{(N_{REZ}-1)} + T^{C(N_{REZ})_1 Q(N_{REZ})} Q^{(N_{REZ})} + Z^{C(N_{REZ})_1}. \quad (115)$$

Substitution of (111) and (115) into (103) gives

$$T^{(Q_{N_{REZ}}-1)} Q^{(N_{REZ}-1)} + T^{(Q_{N_{REZ}})} Q^{(N_{REZ})} = Z^{Q(N_{REZ})}, \quad (116)$$

where

$$\begin{aligned}
T^{(Q_{N_{REZ}}-1)} &= -T^{r((N_{REZ})', N_{REZ})} T^{C(N_{REZ})_1 Q(N_{REZ}-1)}, \\
T^{(Q_{N_{REZ}})} &= T^{z(N_{REZ})} + T^{r((N_{REZ}+1)', N_{REZ})} T^{C(N_{REZ})_2 Q(N_{REZ})} - \\
&\quad T^{r((N_{REZ})', N_{REZ})} T^{C(N_{REZ})_1 Q(N_{REZ})}, \\
Z^{Q_{N_{REZ}}} &= Z^{(N_{REZ})} + T^{r((N_{REZ})', N_{REZ})} Z^{C_{N_{REZ}}}.
\end{aligned} \quad (117)$$

We have obtained a system of equations in which the only vectors of the coefficients of the expansion of the longitudinal field at the centers of the resonators are unknown

$$\begin{aligned}
T^{(Q_1)} Q^{(1)} + T^{(Q_2)} Q^{(2)} &= Z^{Q_1}, \\
T^{(k)} Q^{(k)} &= T^{+(k)} Q^{(k+1)} + T^{-(k)} Q^{(k-1)} + Z^{Q_k}, \quad k = 2, ..., N_{REZ} - 1, \\
T^{(Q_{N_{REZ}}-1)} Q^{(N_{REZ}-1)} + T^{(Q_{N_{REZ}})} Q^{(N_{REZ})} &= Z^{Q_{N_{REZ}}}.
\end{aligned} \quad (118)$$

In our models we have to make several choices: the kind and the number of the basic functions $\varphi_n^{(r)}$, $\varphi_n^{(z)}$ and the testing functions $\psi_n$, and the upper limit of summation in the sums for calculation of matrix elements $T_{s,s'}$.

Summation over $m$ in sums is determined only by computational capabilities.

We used Bessel functions as the testing functions $\psi_s(x) = J_1(\lambda_s x)$, $J_0(\lambda_s) = 0$, then

$$R_{s',s}^{\psi(k',k)} = \int_0^1 \psi_{s'}(x) J_1(b_{k'} \lambda_s x / b_k) x dx = \begin{cases} -\dfrac{b_{k'} \lambda_s}{b_k} \dfrac{J_0\left(\frac{b_{k'} \lambda_s}{b_k}\right) J_1(\lambda_{s'})}{\left(\frac{b_{k'} \lambda_s}{b_k}\right)^2 - (\lambda_{s'})^2}, \\ \dfrac{1}{2} J_1^2(\lambda_s), \quad b_{k'} \lambda_s = b_k \lambda_{s'}, \end{cases} \quad (119)$$

For $z$- basis functions $\varphi_s^{(z)} = J_0(\lambda_s x)$, we have

$$R_{s',s}^{\varphi,z} = \int_0^1 \varphi_s^{(z)}(x) J_0(\lambda_{s'} x) x dx = \delta_{s,s'} \frac{1}{2} J_1^2(\lambda_{s'}) \quad (120)$$

Two sets were used as basis functions $\varphi_s^{(r)}$: A) non-singular Bessel functions $\varphi_s^{(r)}(x) = J_1(\lambda_s x)$. For such set of functions $R_{m,s}^{\varphi,r(k',k)} = R_{m,s}^{\psi(k',k)}$. B) the complete set of functions with singularities

$$\varphi_s^{(r)}(x) = 2\sqrt{\pi} \frac{\Gamma(s+1)}{\Gamma(s-0.5)} \frac{1}{\sqrt{1-x^2}} P_{2s-1}^{-1}\left(\sqrt{1-x^2}\right), \quad (121)$$

where $P_n^m(x)$ are Legendre functions (or spherical functions) of the first kind. For such set of functions [18]

$$R_{s',s}^{\varphi,r(k',k)} = \int_0^1 \varphi_s^{(r)}(x) J_1(b_{k'}\lambda_{s'}x/b_k) x dx = j_{2s-1}\left(\frac{b_{k'}\lambda_{s'}}{b_k}\right). \quad (122)$$

Numerical studies have shown that both types of basis functions give approximately the same results. Functions (121) have singular behavior of the type $\varphi_s^{(r)}(x) \sim (1-x^2)^{-1/2}$ while the real field has behavior of the type $\varphi_s^{(r)}(x) \sim (1-x^2)^{-1/3}$. There are other sets of basis functions that have an arbitrary type of singular behavior $\varphi_s^{(r)}(x) \sim (1-x^2)^{\mu-1}$(see, for example, [19, 20], where it is proposed to use Jacobis hypergeometric polynomials as a complete set of functions), but their numerical implementation is difficult.

Knowing the vectors $Q^{(k)}$, we can calculate the vectors that determine the transverse electric fields at the interfaces between waveguides (see (88),(91),(104),(105),(111),(112), and then find the electric and magnetic fields using the formulas (45),(46),(47). The calculation of the $B_{m,0}^{(q,i)}$ amplitudes is simplified by using the (51) formulas for all subdomains with appropriate subsequent changes in notation

$$\begin{aligned}
B_{m,0}^{(q,1)} &= -\tilde{B}_{m,0}^{(q,1)} + \frac{1}{sh\left(\gamma_m^{(q)}d_q\right) J_1^2(\lambda_m)} \sum_s C_s^{(q,R)} R_{m,s}^{\varphi,r(q)} - \\
&\exp\left(-\gamma_m^{(k')}d_{k'}\right) \frac{1}{sh\left(\gamma_m^{(k')}d_{k'}\right) J_1^2(\lambda_m)} \sum_s C_s^{(q,L)} R_{m,s}^{\varphi,r(q)} \\
B_{m,0}^{(q,2)} &= \tilde{B}_{m,0}^{(q,1)} - \frac{1}{sh\left(\gamma_m^{(q)}d_q\right) J_1^2(\lambda_m)} \sum_s C_s^{(q,R)} R_{m,s}^{\varphi,r(q)} + \\
&\exp\left(\gamma_m^{(q)}d_q\right) \frac{1}{sh\left(\gamma_m^{(q)}d_q\right) J_1^2(\lambda_m)} \sum_s C_s^{(q,L)} R_{m,s}^{\varphi,r(q)},
\end{aligned} \quad (123)$$

where $R_{m,s}^{\varphi,r(q)} = R_{m,s}^{\varphi,r(k',k')}$ for small cross-sectional volumes and $R_{m,s}^{\varphi,r(q)} = R_{m,s}^{\varphi,r(k',k)}$ for large cross-sectional volumes, $C_s^{(q,L)}$, $C_s^{(q,R)}$ - expansion coefficients of transverse electric fields on the left and right sides of the cylindrical volume.

Power flow across the cross-sectional surface of the q-th cylindrical waveguide with longitudinal coordinate $0 \leqslant \tilde{z} \leqslant d_q$ in the absence of current is equal to

$$\Pi_z^{(q)} = \frac{1}{2}\text{Re}\left(\int_0^{b_q} r dr E_r H_\varphi^*\right) = \Pi_{z,0}^{(w,1)} \times \\
\text{Im}\sum_s \left[\varepsilon^* \frac{\lambda_1^2}{\lambda_s^2} \frac{\gamma_s^{(q)}}{\text{Im}\,\gamma_1^{(w,1)}} \frac{b_q^4}{b_{w,1}^4} \frac{J_1^2(\lambda_s)}{J_1^2(\lambda_1)} \left\{B_{s,0}^{(q,1)} \exp\left(\gamma_s^{(q)}\tilde{z}\right) + B_{s,0}^{(q,2)} \exp\exp\left(-\gamma_s^{(q)}\tilde{z}\right)\right\} \times \\
\left\{B_{s,0}^{(q,1)*} \exp\left(\gamma_s^{(q)*}\tilde{z}\right) - B_{s,0}^{(q,2)*} \exp\left(-\gamma_s^{(q)*}\tilde{z}\right)\right\},
\right] \quad (124)$$

$\Pi_{z,0}^{(w,1)}$ is the power carried by the incident wave through the left semi-infinite cylindrical waveguide

$$\Pi_{z,0}^{(w,1)} = \sqrt{1 - \left(\frac{c\lambda_1}{\omega\,b_{w,1}}\right)^2} \frac{\omega^2 b_{w,1}^2 b_{w,1}^2}{4c^2\lambda_1^2 Z_0} J_1^2(\lambda_1) E_0^2, \quad (125)$$

where $E_0 = 1\ V/m, Z_0 = (\mu_0/\varepsilon_0)^{1/2}$.

# 7 Conclusions

The presented approach to the description of inhomogeneous disk-loaded waveguides can be a useful tool in studying the properties of slow wave system. Proposed modification of the coupled integral equations method makes it possible to deal directly with a longitudinal electric field and to obtain approximate equations for the case of a slow change in the waveguide parameters.